\documentclass[%
 reprint,
 amsmath,amssymb,
]{revtex4-2}

\usepackage{graphicx}% Include figure files
\usepackage{dcolumn}% Align table columns on decimal point
\usepackage{bm}% bold math
\usepackage{xcolor}
\usepackage{verbatim}  % Add this in the preamble
\newcommand{\rt}{\right}
\newcommand{\lf}{\left}
\renewcommand\Re{\operatorname{Re}}

\newcommand{\ra}{\rangle}

\newcommand{\s}{\text{s}}

\renewcommand{\i}{\text{in}}

\newcommand{\br}{\mathbf r}

\newcommand{\eq}[1]{Eq.~(\ref{#1})}

\begin{document}

\preprint{APS/123-QED}

\title{Attosecond momentum-resolved resonant inelastic x-ray scattering for imaging coupled electron-hole dynamics}
%\\Tracking Electron-Hole Dynamics in Molecules}% Force line breaks with \\

\author{Maksim Radionov\textsuperscript{1,2}}
\author{Daria Popova-Gorelova\textsuperscript{1,2}}%
\affiliation{%
\textsuperscript{1} Institute of Physics, Brandenburg University of Technology Cottbus-Senftenberg,
Erich-Weinert-Straße 1, 03046 Cottbus, Germany
}%
\affiliation{%
\textsuperscript{2} I. Institute for Theoretical Physics and Centre for Free-Electron Laser Science,
Universität Hamburg, Notkestr. 9, 22607 Hamburg, Germany
}%


%\affiliation{
%\affiliation{
%\affiliation{%


\date{\today}% It is always \today, today,

\begin{abstract}
Improving our understanding of electron dynamics is essential for advancing energy transfer, optoelectronics, light harvesting systems and quantum computing. Recent developments in attosecond x-ray sources provide the fundamental possibility of observing these dynamics with atomic-scale resolution. However, connecting a time-resolved signal to dynamics is challenging due to the broad bandwidth of an attosecond probe pulse. This makes exploring the capabilities of different attosecond imaging techniques crucial. Here, we propose attosecond momentum-resolved resonant inelastic x-ray scattering as a prominent technique for tracking ultrafast dynamics. We demonstrate that the scattering signal contains an information about the instantaneous distribution of charge density across the scattering atoms. To illustrate this, we consider scattering from an $\alpha$-sexithiophene molecule, in which coupled electron-hole dynamics are excited.

%Understanding ultrafast charge dynamics is essential for improving energy transfer, optoelectronics, light harvesting systems, and quantum computing. Attosecond momentum-resolved resonant inelastic X-ray scattering (RIXS) is capable to probe charge dynamics in molecules with attosecond temporal and atomic spatial resolution. However, the connection between the measured signal and the underlying charge-density distribution has not yet been understood. We combine time-dependent perturbation theory with the configuration interaction method to theoretically analyze the signal's behavior concerning charge density. We introduce three site-sensitivity mechanisms that explain this behavior. Our results suggest that attosecond momentum-resolved RIXS is a powerful tool for tracking ultrafast charge dynamics in molecules.
% The periodicity in momentum space corresponds to interatomic distances. Charge density symmetries manifest in the signal.

%\begin{description}
%    \item[Usage]
%    Secondary publications and information retrieval purposes.
%    \item[Structure]
%    You may use the \texttt{description} environment to structure your abstract;
%    use the optional argument of the \verb+\item+ command to give the category of each item. 
%\end{description}
\end{abstract}

\maketitle

%Current font: \fontname\font

%\tableofcontents

%\the\textwidth

%\the\textheight

%\the\columnwidth

The absorption of an optical or an ultraviolet photon can lead to the neutral excitation of a molecule driving the motion of a charge density \cite{Han2013}. This motion plays a crucial role in energy transfer \cite{Vismarra2023}, optoelectronics \cite{Anantharaman2021, Fiori2013, Hong2014}, photoelectrochemical sensing \cite{Mohanta2024, Yan2025, Li2025, Wang2023, Liu2025}, quantum computing \cite{Andergassen2010, Han2022, Syurakshin2024, Hastrup2020, Scheidegger2025}, and for light-harvesting systems \cite{Kundu2017,Ostroverkhova2016}. The time scales of electron dynamics range from sub- to few femtoseconds, and the relevant spatial scales are interatomic distances ranging from \r Angstrom to nanometers. The direct observation of electron dynamics requires a combination of attosecond temporal and atomic spatial resolution \cite{Goulielmakis2010, Foehlisch2005}.

Attosecond pump--probe spectroscopy has matured over the past two decades providing the temporal resolution necessary to observe electron dynamics using spectroscopy techniques \cite{Agostini2001, Sansone2006, Sansone2010, Ferrari2010, Calegari2014, Agostini2024}. Attosecond imaging requires going beyond spectroscopy and the application of scattering and diffraction techniques, which has recently become possible at free-electron laser sources \cite{Duris2020, Yan2024, Prat2023, Ulmer2025, Kuschel2025}. Attosecond imaging using a pump-probe scheme remains challenging, but efforts are being made to improve the control and stability of attosecond x-ray experiments \cite{Maroju2023, Guo2024, Shuai2024}. Attosecond imaging poses an additional challenge in interpreting signals due to the broad bandwidth of attosecond pulses and the interaction of light with non-stationary electron states. In order to guide experimental developments, it is necessary to investigate and propose beneficial schemes for imaging electronic motion using broad-bandwidth pulses. 

X-ray scattering (XRS) with hard x rays provides \linebreak(sub-)nanometer spatial resolution. This process is governed by two terms of the light-matter interaction Hamiltonian, namely, $\mathbf A\cdot \mathbf p$ and $\mathbf A^2$, where $\mathbf A$ is the vector potential of an x-ray field and $\mathbf p$ is the momentum operator \cite{Santra2008}. If an x-ray pulse is resonant with the core excitation energy of a system, the former term dominates, and x-ray scattering is referred to as resonant  \cite{Fink2013}. If the x-ray pulse is detuned from any transition, the latter term dominates, and scattering is referred to as non-resonant. Non-resonant XRS \cite{Yong2021, Dixit2012, Simmermacher2019} and sum frequency diffraction \cite{Rouxel2018} have been proposed to follow valence-electron motion, attosecond ring currents \cite{Carrascosa2021}, and conical intersection dynamics \cite{Keefer2021}. Hybrid x-ray/electron diffraction schemes have also been suggested to trace coupled electron-nuclear dynamics \cite{Yong2022, Yong2023}.

%The recent advancements of intense attosecond x-ray free-electron lasers \cite{Emma2010, Osipov2018, Walter2022, Dragone2013, Ferguson2015, Struder2010, Duris2020, Huang2016, Franz2022, Guo2024, Yan2024, Ilchen2025} have enabled the first attosecond x-ray imaging experiments at the Linac Coherent Light Source \cite{Ulmer2025, Kuschel2025}, offering \AA ngstrom spatial resolution. However, attosecond temporal resolution has not yet been combined with such spatial resolution.  Existing implementations of x-ray diffraction (XRD) \cite{Kupitz2014, Glownia2016,Yong2020} and photoelectron momentum microscopy \cite{Kutnyakhov2020, Baumgartner2022, Schiller2025, Keunecke2020, Bange2023, Schmitt2022, Wallauer2021} in the pump-probe scheme remain limited to the femtosecond scale.

The advantage of ultrafast resonant x-ray scattering over non-resonant XRS for attosecond imaging is that it enables the selective enhancement of the scattering contribution from the (quasi)particles involved in the dynamics \cite{Daria2015, Daria2015-2}. The conventional resonant x-ray scattering technique measures the momentum of elastically scattered photons \cite{Fink2013}, while resonant inelastic x-ray scattering (RIXS) is a spectroscopic technique \cite{Ament2011}. Momentum-resolved RIXS combines the strengths of both techniques. Several theoretical works have been developed to describe RIXS \cite{Chen2019, Freibert2024-1, Freibert2024-2} and momentum-resolved RIXS \cite{Daria2015, Daria2015-2,  Fouda2021, Venkatesh2025} from non-stationary electron systems. It has been demonstrated that an attosecond momentum-resolved resonant x-ray scattering signal is non-centrosymmetric due to microscopic electron currents and cannot be straightforwardly related to the time-dependent electron density \cite{Daria2015, Daria2015-2}. It has been suggested that the centrosymmetric component of the signal correlates with the electron density \cite{Daria2018}. However, this idea has not been elaborated upon further. In this study, we demonstrate how to observe the density of a coupled electron-hole using attosecond momentum-resolved RIXS.

\begin{figure*}[tb]
\includegraphics[width=1.0\textwidth]{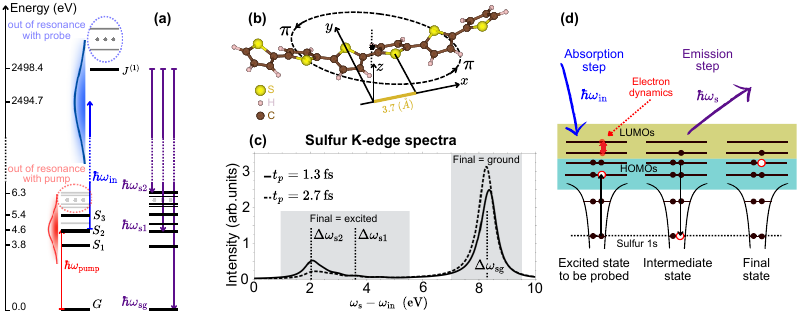}% Here is how to import EPS art
\caption{\label{fig:SexithiopheneSpectra} (a) Illustration of attosecond pump-probe experiment. (b) Sexithiophene molecule. (c) Sulfur K-edge spectra of an excited sexithiophene molecule at different time delays ($\mathbf{k}_{\text{in}}||y$, $\boldsymbol{\epsilon}_\text{in}||x$). (d) Illustration of states involved in the dynamics and transitions.}
\end{figure*}

We describe an experiment, in which a pump pulse excited a molecule into a coherent superposition of the excited singlet states $|\Psi_{n}\ra$ with a hole in HOMO orbitals and an electron in LUMO orbitals  with corresponding eigenenergies $\varepsilon_n$ creating an excited state $\Psi(t) = \sum_{n\ge1}C_n e^{-i\varepsilon_n t}|\Psi_n\ra$. Modern experimental capabilities make it possible to create such a coherent superposition of molecular excited states \cite{Calegari2014, Trabattoni2019, Worner2022, Calegari2015, Lara-Astiaso2018, Kraus2015, Mansson2021, Belshaw2013}. An x-ray probe pulse acts on the molecule after a nonzero time delay $t_p$ and does not temporally overlap with the pump pulse. 
%The total Hamiltonian of the electronic system and the probe x-ray pulse is
%\begin{align}
%\hat H = \hat H_0+\frac{1}{c}\hat{\mathbf A}\cdot\hat{\mathbf p}+\hat H_r,
%\end{align}
% $\hat{\mathbf A}$ is the vector potential of the x-ray field, $\hat{\mathbf p}$ is the momentum operator 
%where $\hat H_0$ and $\hat H_r$ is the Hamitonians of the electronic system and the x-ray field, respectively, and $c$ is the speed of light. 
We consider the process in which an x-ray probe pulse is resonant to a transition of a core electron to energy levels below the Fermi level (see Fig.~\ref{fig:SexithiopheneSpectra}(a)). As these levels are occupied in the ground state, the scattering signal is primarily caused by excited molecules. This results in a substantial benefit of this method being unaffected by the background due to stationary electrons. We further assume that the x-ray pulse has a Gaussian-shaped temporal profile with duration $\tau_p$ defined as the full width at the half maximum and has a bandwidth that is considerably larger than the energy splitting between the excited states involved in the dynamics, but smaller than the energy difference between the ground and the excited states. The first condition is necessary to have a sufficient temporal resolution to resolve the excited-state dynamics. The latter condition is beneficial for the exclusion of the signal due to the interference with the ground state. The interference can appear, because the total wave function $|\Psi^{\text{(tot)}}\ra = C_0e^{-i\varepsilon_0 t_p}|\Psi_0\ra+\sqrt{1-|C_0|^2}|\Psi(t)\ra$ includes a contribution from the ground state $|\Psi_0\ra$. $C_0$ is the corresponding complex expansion coefficient. Our assumption describes the experimental conditions, when a time-resolved signal is sensitive only to the excited-state dynamics and would not be distracted by the interference with the ground state.
%The total scattering probability is then given by the product of the probability that the molecule is in an excited state and the scattering probability from an excited molecule (see {\blue Appendix}). 
In Appendix \ref{app:EquationDerivation}, we derive the scattering probability with these assumptions
\begin{widetext}
\begin{eqnarray}
P(\mathbf Q,t_p) = P_{\text{e}}P_0\, \theta(\mathbf n_{Q}) \sum\limits_{F} \frac{1}{\omega_\text{s}} \left|\sum\limits_n C_n e^{-i\varepsilon_nt_p} e^{-\frac{\left(\varepsilon_F+\omega_\text{s}-\varepsilon_n-\omega_{\text{in}}\right)^2\tau_p^2}{8\ln{2}}} \sum\limits_J e^{i\mathbf{Q}\cdot \mathbf{R}_J} \cdot \frac{\langle\Psi_F|\boldsymbol{\epsilon}_\text{s}^* \cdot \boldsymbol{\nabla}|\Psi_J\rangle \langle\Psi_J|\boldsymbol{\epsilon}_{\text{in}} \cdot \boldsymbol{\nabla}|\Psi_n\rangle}{\left(\omega_\text{s}+\varepsilon_F-\varepsilon_J+i\frac{\Gamma}{2}\right)} \right|^2,
\label{eq:P_k_F}
\end{eqnarray}
\end{widetext}
where $P_{\text{e}} = 1-|C_0|^2$ is the probability that the molecule is in an excited state and $P_0$ is a constant prefactor.  $\mathbf Q = \mathbf k_\text{s}-\mathbf k_\i$ is the scattered vector; $\mathbf{k}_{\text{in}}$, $\omega_{\text{\text{in}}}$, and $\mathbf{k}_{\s}$, $\omega_{\s}$ denote the wave vector and photon energy of incoming and scattered radiation, respectively. $\boldsymbol{\epsilon}_{\text{in}}$ is the polarization of the incoming pulse. Since photons are scattered in various possible directions, they have different polarizations, which leads to an additional dependence of the signal on the direction of the scattered vector $\mathbf n_{Q}$ through the emission transition matrix elements. This dependence does not carry relevant information. We factored it out into $\theta(\mathbf n_{Q})$ by the application of a reflective polarizer for the polarization $\boldsymbol{\epsilon}_{\text{s}}$, please see Ref.~\cite{Daria2015-2} for details. $|\Psi_J\ra$ is the intermediate state with a hole at a localized core orbital at position $\mathbf R_J$. Its energy is denoted as $\varepsilon_J$ and its lifetime broadening as $\Gamma$. The first sum runs over all possible final states $|\Psi_F\ra$ with energies $\varepsilon_F$. In the following analysis, we focus on such transitions that result in a scattered photon being close to the incoming photon energy (see Fig.~\ref{fig:SexithiopheneSpectra}(a)). In this case, a final state would either coincide with the ground state, one of states involved in the dynamics or some other valence-excited state. Any possible transition is inelastic, since the initial state is a nonstationary state. We use atomic units for this and the following equations.

We demonstrate the power of attosecond resonant inelastic x-ray scattering to reveal information about coupled electron-hole dynamics by considering scattering on the sexithiophene molecule shown in Fig.~\ref{fig:SexithiopheneSpectra}(b). Sexithiophene is noteworthy for optoelectronic applications due to its special optical properties  governed by excitonic excited states \cite{Barbarella1999, Stallinga2004, Mason2011}. We take into account many-body effects due to electron-hole coupling using the restricted active space configuration interaction (RASCI) method \cite{Roos1992} implemented in the MOLCAS package \cite{Aquilante2016}. We calculate eigenstates and eigenenergies of sexithiophene, and use them to calculate the electron density and the scattering signal with \eq{eq:P_k_F} (see Appendix \ref{app:CompDetails} for further computational details). 

\begin{figure*}
\includegraphics[width=1.0\textwidth]{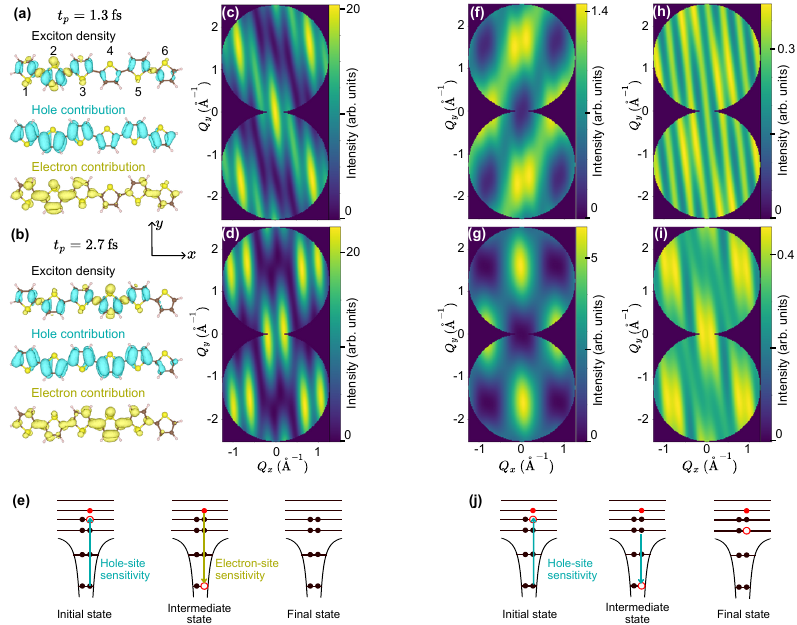} %[width=\textwidth]
\caption{\label{fig:MomMaps} (a) - (b) Exciton density, hole and electron contributions to the exciton density at (a) $t_p=1.3$ fs and (b) $t_p=2.7$ fs visualized with the VESTA package \cite{Momma2011}. Yellow and cyan isosurfaces correspond to the negatively- and positively-charged regions. (c)--(d) The even part of the momentum maps at $Q_z = 0$, $\lf[P(Q_x, Q_y, 0)+P(-Q_x, -Q_y, 0)\rt]/(2\theta(\mathbf n_Q))$, with the final state being the ground state at (c) $t_p=1.3$ fs and (d) $t_p=2.7$ fs and (e) illustration of the involved transitions. (f)--(i) The same as (c)--(d), but with the final state being a valence-excited state at (f) $\omega_\s-\omega_\i=\Delta\omega_{\s2}$ and $t_p=1.3$ fs; (g) $\omega_\s-\omega_\i=\Delta\omega_{\s2}$ and $t_p=2.7$ fs; (h) $\omega_\s-\omega_\i=\Delta\omega_{\s1}$ and $t_p=1.3$ fs; and (i) $\omega_\s-\omega_\i=\Delta\omega_{\s1}$ and $t_p=2.7$ fs. (j) Illustration of the scattering process with a final state being a valence-excited state.}
\end{figure*}

We assume that a pump pulse created a coherent superposition of the first three bright excited singlet states with the energies $3.8$ eV, $4.6$ eV and $5.4$ eV at time $t_p = 0$. We set the ground-state energy to zero. Excited states with higher energies can be excluded from the superposition in the pump process, since they are energetically separated (see Fig.~\ref{fig:SexithiopheneSpectra}(a)). We select the coefficients $C_1$, $C_2$ and $C_3$ to be equal to $1/\sqrt{3}$, which can be achieved by selecting the appropriate pump-pulse parameters. The conclusions of our study are independent of the specific choice of coefficients. 
We show the difference between $\rho(\br,t)$ and the ground-state electron density at two different time delays in Figs.~\ref{fig:MomMaps} (a) and (b). We also disentangle the hole and electron contributions to the differential electron density as described in Appendix \ref{app:EquationDerivation} and show them below it. 
%{\blue As the bright excited states can be reached via dipole-allowed transitions, the exciton density exhibits the same twofold rotational symmetry as the ground-state density.} \textcolor{olive}{Mb link somewhere? Very untrivial statement and depends on $\epsilon_\text{in}$}

We assume that the x-ray probe pulse has a mean photon energy $\omega_{\text{in}}=2490$ eV and a duration $\tau_p=300$ as, which corresponds to a bandwidth of 2.6 eV, $\boldsymbol{k}_{\text{in}}||y$ and $\boldsymbol \epsilon_\i || x$. Such an x-ray pulse is resonant with transitions from the states $|\Psi_{1,2,3}\ra$ to the intermediate states with a hole in the sulfur $1s$ orbitals. The energies of the core-excited states of sexithiophene vary slightly depending on the location of the core hole, by no more than 120 meV. This forms groups of core-excited states, depending on the character of the excitations in the LUMO orbitals. We set $\omega_{\text{in}}$ such that transitions to the lowest-energy group of core-excited states dominate, thus facilitating the analysis. The lifetime broadening of sulfur $1s$-excited states $\Gamma$ is 0.59 eV \cite{Krause1979}.

Figure~\ref{fig:SexithiopheneSpectra}(c) shows momentum-unresolved spectra at two different time delays. The right intensive peak corresponds to emission into a final state that is the ground state. The peak is actually composed of several peaks. The positions of the individual peaks do not vary over time, but the intensity does, which leads to the illusion that the peak shifts. The left broad peak is due to emission with final states being valence excited states. The bandwidth of the individual peaks is determined by both the spectral bandwidth of the probe pulse and the Lorenzian lifetime broadening.

Figures \ref{fig:MomMaps} (c), (d), (f)-(i) show the centrosymmetric part of the time- and momentum-resolved RIXS signal, $\lf[P(\mathbf Q)+P(-\mathbf Q)\rt]/2$, at the three different scattered energies outlined in Fig.~\ref{fig:SexithiopheneSpectra} (c), and at different time delays. This part of the signal follows the electron density, since it involves the same interference terms $\sum_{n_1,n_2>n_1}\Re(C^*_{n_2}C_{n_1}e^{-i(\varepsilon_{n_1}-\varepsilon_{n_2})t_p} )$ as the density does \cite{Daria2018}. Here, we apply the polarizer for $\boldsymbol\epsilon_\s||z$ and divide the signal by $\theta(\mathbf n_Q)$. The total time- and momentum-resolved RIXS signal shown in the supplementary Fig.~S1 is not centrosymmetric due to the presence of currents \cite{Daria2015}. Since the intermediate states have localized holes on sulfur atoms, sulfur atoms are the scattering atoms. The $\mathbf Q$-dependent part of the signal is a combination of periodic functions with periods $1/\lambda_{ik}$, where $\lambda_{ik}  = |\mathbf R_{i}-\mathbf R_{k}|/2\pi$ with $i$ and $k$ denoting different sulfur atoms. The spatial resolution is sufficient to resolve oscillations due to nearest-neighbour sulfur atoms separated by 4.3 \r A.

%Accessible momenta maps are limited by the momentum conservation law leading to $-\omega_\i/c\le Q_{x,y}\le\omega_\i/c$. \textcolor{olive}{It is for x, for y it is bigger twice} Therefore, the spatial resolution is half of the x-ray wavelength of 5 \r A and is sufficient to resolve oscillations due to nearest-neighbour sulfur atoms separated by 4.3 \r A.

%Absorption destroys the coherent superposition, but the related transition amplitude is determined by the initial nonstationary state. Transition amplitude of emission does not carry information about the initial state. But since scattering involve the product of the two amplitudes, the signal contains the information about the electron dynamics at the time of measurement.

The scattering signal is sensitive to the excited-state dynamics through the time-dependent absorption amplitudes. During the absorption step of the scattering process, a core electron localized on a sulfur atom $i$ fills the delocalized hole. This transition is only possible if the hole density around the atom $i$ at the time of the measurement is considerable. Interference fringes in the scattering signal appear, if a pair of atoms scatters. Thus, maxima separated by $1/\lambda_{ik}$ in the signal indicate that the hole density is simultaneously non-zero around a pair of atoms $i$ and $k$. For example, the hole density on the most widely separated atoms 1 and 6 is considerable at time $t_p = 1.3$ fs (see Fig.~\ref{fig:MomMaps} (a)). In contrast, the hole density on the atom 6 is negligible at time $t_p=2.7$ fs (see Fig. \ref{fig:MomMaps} (b)). Consequently, the shortest-period oscillations can clearly be observed in all momentum maps at $t_p = 1.3$ fs, but are not visible in the maps at $t_p = 2.7$ fs. The scattering signal, thus, encodes information about the hole density at the time of measurement. 

%The signal is also sensitive to the electron contribution to the excited-state dynamics, but in a less trivial way. We explain this connection using an independent-electron picture although our calculations include many-body effects. If we considered states $|\Psi_1\ra$ and $|\Psi_2\ra$ in the independent-particle picture, they would have an electron in a LUMO orbital. A transition to an intermediate state $|\Psi_J\ra$ would be possible, if it also has an electron in the same LUMO orbital, see {\blue Fig}. Scattering with the final state being the ground state is then due to the transition of the LUMO electron into a vacant core state. A $\mathbf Q$-dependence in the signal is then due to pairs of atoms that have a contribution from the LUMO orbital around them.  

It is also relevant to investigate the electron contribution to the excited-state dynamics. This contribution is due to the excited states having electrons distributed in the LUMO states. It turns out that the signal is also sensitive to this electron contribution, if the molecule's final state is the ground state. This connection is non-trivial, so let us  explain it using the independent-particle picture. After the action of the probe pulse, a core electron is excited to a HOMO orbital, while the same LUMO orbital remains occupied after absorption (see Fig.~\ref{fig:MomMaps}(e)). During emission, the LUMO electron fills the core hole, placing the system in the ground state. For emission to be possible, a LUMO distribution on an atom must be considerable. Therefore, both an excited-state hole and electron distribution must be considerable around a scattering atom for scattering to be possible. A $\lambda_{ik}$ oscillation in the corresponding signal reflects the simultaneous presence of an electron-hole pair on atoms $i$ and $k$. The hole and electron distributions of optically-excited sexithiophene move almost synchronously, enhancing the contrast of the oscillations. 

%{\violet The signal clearly shows high-$\lambda_{ik}$ oscillations at $t_p=T/2$\textcolor{olive}{1.3 fs} due to the exciton density being concentrated at the edges of sexithiophene and low-$\lambda_{ik}$ oscillations due to the exciton density being localized in the middle at $t_p=T$\textcolor{olive}{2.7 fs}.}

%%%%Beyond the independent particle picture, we should note that the overlap between the electron distributions in the LUMO orbitals of the intermediate states and those of all states $|\Psi_{1,2,3}\ra$ should be non-zero. Strictly speaking, the signal is sensitive to the projection of the electron contribution onto the LUMO orbitals of the intermediate states. Figures~\ref{fig:MomMaps}(c) and ~\ref{fig:MomMaps}(d) show the corresponding signal. 

% If the final state is the ground state, a vacant core state is filled by the electron from LUMO states of $|\Psi_J\ra$ (see {\blue Fig.}). If this electron is distributed over a pair of atoms $i$ and $k$ simultaneously, then the $2\pi/(\mathbf R_{i}-\mathbf R_{k})$-oscillation can appear.

Now, let us consider transitions in which the final state is not the ground state, but rather a valence-excited state. Due to the same absorption step, the signal would still be sensitive to the excited-state hole density. However, the connection to the electron contribution would not hold. We use the independent-particle picture again to illustrate this mechanism (see Fig.~\ref{fig:MomMaps}(j)). For absorption and subsequent emission to a final state being an excited state, all states involved must have the same LUMO orbital being occupied by an electron. This condition excludes some possible final states and the only role of the LUMO electron is now to select possible final states. Emission is due an electron from a HOMO orbital filling the core hole. The structure of this HOMO orbital determines whether the $\mathbf Q$ dependence due to the hole distribution around scattering atoms is suppressed or enhanced as can be observed in Figs.~\ref{fig:MomMaps} (f)-(i). Unlike the energetically separated ground state, the valence-excited states are close in energy, making it impossible to disentangle the role of an individual final state in the signal. Nevertheless, analyzing the signal at such scattering energies is advantageous. For scattering with the final state being the ground state, interference fringes are possible, only if both excited-state electron and hole distributions are considerable around an atomic pair. If the signal is now analyzed at a different scattered energy, new interference fringes can occur. This would indicate that a considerable hole distribution exists around some atoms where the electron distribution is negligible.

%An electron from a core orbital fills the vacancy in HOMO, and an electron occupies the same LUMO orbital in the intermediate state. In the emission step, an electron from some HOMO orbital fills the core hole, and the same LUMO orbital is occupied with an electron. A possible final state must then have a same LUMO orbital to be occupied by an electron  

% The dominating contributions to the momentum maps are due to periodic functions with periods 1.7 \AA$^{-1}$ and 0.3 \AA$^{-1}$ along the $Q_x$ direction. 

We introduced a method to extract the information about the time-dependent charge density evolving due to coupled electron-hole dynamics with the momentum-resolved RIXS. It is a powerful method due to the resonant enhancement of the signal from moving particles, and its orbital and site sensitivity.

\section*{Acknowledgements} This work was funded by the Volkswagen Foundation under Grant No. 96237. The authors acknowledge helpful discussions with Dr. Kalyani Chordiya.
\appendix
\section{\label{app:CompDetails}Computational details}

The calculations of the electronic states of the isolated sexithiophene molecule are performed using the MOLCAS software \cite{Aquilante2016}. The molecular geometry has been taken from a molecular structure database PubChem \cite{Bolton2008}. To calculate the ground state, we use the Hartree-Fock approach \cite{Thijssen2007} within the ANO-S-VTZP atomic orbital basis set \cite{Pierloot1995}.
%\begin{equation}
%    |\phi_p(\boldsymbol{r})\rangle=\sum\limits_{N,o}\Xi_{p,N,o}|\xi_{N,o}(\boldsymbol{r})\rangle,\label{eq:MolecularOrbital}
%\end{equation}
%where $o$ denotes the index of the basis atomic orbital of the atom $N$; $p$ is the molecular orbital index; $|\xi(\boldsymbol{r})\rangle$, $\Xi$ are the basis atomic orbital and its coefficient.
We use these molecular orbitals without further optimization to obtain the excited states within the restricted active space configuration interaction (RASCI) method \cite{PerAke1990}. The RASCI method represents the excited state as a sum of configuration state functions (CSFs) \cite{Fales2020}. We restrict the expansion to the singly excited CSFs, such that the molecular eigenstate is $|\Psi_n\rangle = \sum\limits_{i,a} c^{(n)}_{ia} |\Phi_i^a\rangle$, where $|\Phi_i^a\rangle$ denotes a CSF with a hole in the $i^\text{th}$ molecular orbital and an additional electron on the $a^\text{th}$ molecular orbital. The calculations of the valence excited states converge with respect to the active space on 100 inactive orbitals, 27 RAS1 orbitals, and 27 RAS3 orbitals \cite{Sauri2011}. Calculating charge dynamics, we neglect the nuclear motion assuming that the coherence loss is negligible for the considered time scales of few femtoseconds \cite{Despre2015, Despre2018, Matselyukh2022, Grell2023}.

We have compared our results with the available experimental data. In our calculations, the energy difference between the first excited state and the ground state is 3.8 eV. There are no data on the isolated molecules, but there are many experiments on sexithiophene in the solid phase and in various solutions. The first absorption peak appears at approximately 2.3 -- 3.0 eV \cite{Yassar1995, Horowitz1997, Muccini1998, Moller2001, Zhai2010, Zhao2021}.%Experimentally, the first absorption peak in the $\alpha$-crystal phase appears at approximately 2.3 eV \cite{Horowitz1997, Muccini1998, Moller2001}, while in the tetrahydrofuran solution, the corresponding peak is observed at approximately 3.0 eV \cite{Zhai2010}, and in the dioxane \cite{Yassar1995} and dimethylformamide \cite{Zhao2021} solutions at 2.8 eV.

%\begin{figure}[t]
%\includegraphics[width=1.0\columnwidth]{RASCI-calculation.pdf}% Here is how to import EPS art
%\caption{\label{fig:RASCI}Restricted active spaces for calculations of the valence-excited and 1$s$-excited states.}
%\end{figure}

We calculate the core-excited intermediate states using the RASCI with the highly excited states (HEXS) method, following the procedure described in \cite{Nenov2019}.  The first six molecular orbitals are predominantly represented as linear combinations of six sulfur 1$s$ basis atomic orbitals. In six different calculations, we put one of these molecular orbitals into the RAS1 subspace. The RAS3 subspace is kept the same as for the valence-excited states (see the supplementary Fig.~S2). Since the core molecular orbitals are delocalized, the core hole in the calculated excited state, $J'$, is also delocalized. The first six core-excited states are nearly degenerate in energy (the energy difference is less than 120 meV). This means that the states $J'$ can be transformed into six states $J$ with a localized core hole. The state $J'$ can then be expressed as a linear combination of the CSFs with a hole in the sulfur 1$s$ molecular orbital $j'$ and an electron in an unoccupied orbital $b$: $|\Psi'_{J'}\rangle=\sum\limits_{b}c^{(J')}_{b}|\Phi_{j'}^b\rangle$. 

We express the field operators $\hat{\psi}$ and $\hat{\psi}^\dagger$ in the basis of the electron creation $\hat{c}^\dagger$ and annihilation $\hat{c}$ operators and rewrite the transition matrix elements between the valence and sulfur 1$s$-excited states in the basis of the CSFs:
\begin{align}
\langle\Psi'_{J'}|\hat{\psi}^\dagger e^{i\boldsymbol{k}\cdot\boldsymbol{r}}(\boldsymbol{\epsilon}\cdot\boldsymbol{\nabla})\hat{\psi}|\Psi_n\rangle = \sum\limits_{p,q}\sum\limits_{a,i,b}{c_b^{(J')}}^*c_{ia}^{(n)}\nonumber\\
\times\langle\Phi_{j'}^b|\hat{c}_p^\dagger \hat{c}_q|\Phi_i^a\rangle\langle\phi_p|e^{i\boldsymbol{k}\cdot\boldsymbol{r}}\boldsymbol{\epsilon}\cdot\boldsymbol{\nabla}|\phi_q\rangle,\label{eq:n-J-matrix}
\end{align}
where the matrix element $\langle\Phi_{j'}^b|\hat{c}_p^\dagger \hat{c}_q|\Phi_i^a\rangle$ is nonzero only if the two configurations differ by at most one occupied orbital. Since the molecular orbital $\phi_{j'}$ corresponds to the sulfur $1s$ orbital, which is doubly occupied in the configuration $|\Phi_i^a\rangle$ and singly occupied in $|\Phi_{j'}^b\rangle$, the annihilation operator $\hat{c}_q$ yields a nonzero value only when $q = j'$. We expand the molecular orbital as a linear combination of the basis atomic orbitals, $|\phi_{j'}\rangle=\sum\limits_{N,o} \Xi_{j',N,o}|\xi_{N,o}\rangle$. Here, $o$ denotes the index of the basis atomic orbital of the atom $N$. For the first six molecular orbitals $j'$, the coefficients $\Xi_{j',N,o}$ are significant only for strongly localized sulfur $1s$ atomic orbitals. Assuming that the x-ray wavelength is much larger than the spatial extend of the $1s$-orbital leads to the approximation $e^{i\textbf{k}\cdot\textbf{r}}\Xi_{j',N,o}|\xi_{N,o}\rangle \approx e^{i\textbf{k}\cdot\textbf{R}_N}\Xi_{j',N,o}|\xi_{N,o}\rangle$. The transition matrix element can then be expressed then as
\begin{align}
\langle\phi_p|e^{i\boldsymbol{k}\cdot\boldsymbol{r}}(\boldsymbol{\epsilon}\cdot\boldsymbol{\nabla})|\phi_{j'}\rangle&= \sum\limits_{N,o}\Xi_{j',N,o}\langle\phi_p|e^{i\boldsymbol{k}\cdot\boldsymbol{r}}(\boldsymbol{\epsilon}\cdot\boldsymbol{\nabla})|\xi_{N,o}\rangle\nonumber\\
&\approx \sum\limits_{N,o}\Xi_{j',N,o}e^{i\boldsymbol{k}\cdot\boldsymbol{R}_N}\langle\phi_p|\boldsymbol{\epsilon}\cdot\boldsymbol{\nabla}|\xi_{N,o}\rangle.\label{eq:mol-mol-matrix}
\end{align}

\section{\label{app:EquationDerivation}Equation derivation}

The general expression for attosecond momentum-resolved RIXS has been derived in Ref.~\cite{Daria2015} using the time-dependent second-order perturbation theory and the second quantization formalism. In the present work, we follow the same derivation steps, but (i) do not integrate the signal over the energy window and (ii) do not apply the mean energy approximation for the non-stationary electronic system.

We first show that the scattering probability does not have a background due to the ground state. If the initial state is $|\Psi^{(\text{tot})}\rangle$, the scattering probability can be expressed as a function of the probe pulse parameters $\omega_\text{in}$ and $\tau_p$ \cite{Daria2015}:
\begin{align}
    P(\omega_{\text{in}}, \tau_p)=a\sum\limits_{F,s_\text{s}}\left|C_0 e^{-\frac{(\varepsilon_F+\omega_\text{s}-\varepsilon_0-\omega_\text{in})^2\tau_p^2}{8\ln{2}}} b_{0,F}\right.\nonumber\\
    \left.+\sum\limits_{n\geq 1}\sqrt{1-|C_0|^2} C_n e^{-\frac{(\varepsilon_F+\omega_\text{s}-\varepsilon_n-\omega_\text{in})^2\tau_p^2}{8\ln{2}}} b_{n,F}\right|^2,\label{eq:P_factor_1}
\end{align}
where coefficients $a$ and $b_{n,F}$ depend neither on $\omega_\text{in}$ nor on $\tau_p$. The first term in the modulus is the contribution of the ground state to the scattering signal. Since in our case, the energy difference between the ground state and any excited state is much larger than the energy difference between any two excited states, the factor $(\varepsilon_F+\omega_\text{s}-\varepsilon_0-\omega_\text{in})$ in the exponent in the first term is considerably larger than that in the other terms, $(\varepsilon_F+\omega_\text{s}-\varepsilon_n-\omega_\text{in})$, for any $n\geq 1$. Consequently, the exponent in the first term is much smaller than those in the others. For our parameters $\tau_p=300\text{ as}$, $\omega_\text{in}=2490\text{ eV}$, it is approximately ten times smaller. Therefore, we neglect the first term, and express the scattering probability as:
\begin{align}
    P(\omega_\text{in},\tau_p)\approx aP_\text{e}\sum\limits_{F,s_\text{s}}\left|\sum\limits_{n\geq1}C_n e^{-\frac{(\varepsilon_F+\omega_\text{s}-\varepsilon_n-\omega_\text{in})^2\tau_p^2}{8\ln{2}}} b_{n,F}\right|^2, \label{eq:factrorization}
\end{align}
where $P_\text{e}=\left(1-|C_0|^2\right)$ is the probability that the system is in an excited state. The contribution of the ground state is now in the pre-factor $P_\text{e}$.

Using the approximation in Eq.~(\ref{eq:mol-mol-matrix}) for the transition matrix element in Eq.~(\ref{eq:n-J-matrix}) and the approximation in Eq.~(\ref{eq:factrorization}), we obtain the following expression for the attosecond momentum-resolved RIXS signal:
\begin{widetext}
\begin{align}
P(\mathbf{k}_\text{s}, t_p) =& P_\text{e}\,\theta(\mathbf n_{Q})P_0\sum\limits_F \frac{1}{\omega_\text{s}} \left|\sum\limits_n C_n e^{-i\varepsilon_nt_p}e^{-\frac{\left(\varepsilon_F+\omega_\text{s}-\varepsilon_n-\omega_{\text{in}}\right)^2\tau_p^2}{8\ln{2}}} \sum\limits_{J'} \frac{1}{\left(\omega_\text{s}+\varepsilon_F-\varepsilon_{J'}+i\frac{\Gamma}{2}\right)} \right.\nonumber\\
&\times\left( \boldsymbol{\epsilon}_\text{s}^* \cdot \sum\limits_{p_1, q_1} \sum\limits_{N_1, o_1} e^{-i\boldsymbol{k}_\text{s} \boldsymbol{R}_{N_1}} \Xi_{p_1,N_1,o_1}^* \langle\xi_{N_1,o_1}|\boldsymbol{\nabla}|\phi_{q_1}\rangle\langle\Psi_F|\hat{c}_{p_1}^\dagger \hat{c}_{q_1}|\Psi'_{J'}\rangle  \right) \nonumber\\
&\left. \times \left( \boldsymbol{\epsilon}_{\text{in}} \cdot \sum\limits_{p_2, q_2} \sum\limits_{N_2,o_2} e^{i\boldsymbol{k}_{\text{in}} \boldsymbol{R}_{N_2}} \Xi_{q_2,N_2,o_2} \langle\phi_{p_2}|\boldsymbol{\nabla}|\xi_{N_2,o_2}\rangle \langle \Psi'_{J'} |\hat{c}_{p_2}^\dagger \hat{c}_{q_2}|\Psi_n\rangle \right)\right|^2,
\label{eq:P_k_F_2}
\end{align}
\end{widetext}
where $P_0 = 2\pi^3 I_0 \tau_p^2/(\ln{2} cV\omega_{\text{in}}^2)$ with $I_0$ being the peak intensity of the probe pulse, $c$ - the speed of light, and $V$ - the quantization volume.

As explained in Appendix \ref{app:CompDetails}, we represent the state $|\Psi_{J}\rangle$ with a hole localized on a single sulfur atom $j$ as a linear combination of the calculated intermediate states $|\Psi_{J}\rangle\approx\sum\limits_{j'} \alpha_{jj'}|\Psi'_{J'}\rangle$. Under this approximation, Eq.~(\ref{eq:P_k_F}) can be obtained from Eq.~(\ref{eq:P_k_F_2}). We have checked that the scattering probabilities calculated with and without this assumption differ by less than 1\%.

During the excitation process, the absorption of an UV photon creates an electron-hole pair. Consequently, the charge density of the excited state $\rho=\left|\Psi(t)\right|^2$ differs from that of the ground state $\rho_G$ by one hole in the RAS1 space (HOMOs) and one electron in the RAS3 space (LUMOs). We denote this difference by the exciton density. The exciton density has the contribution due the LUMOs electron density $\rho_\text{e}(t_p)$ and the HOMOs hole density $\rho_\text{h}(t_p)$, which we disentangle with the procedure below:
\begin{widetext}
\begin{align}
    \rho(t_p, \boldsymbol{r})-\rho_{G}(\boldsymbol{r}) &=\rho_\text{e}(t_p)+\rho_\text{h}(t_p)=\sum\limits_{m,n} C_m^* C_n e^{i(\varepsilon_m-\varepsilon_n)t_p} \sum\limits_{p,q}\langle\Psi_m|\hat{c}_p^\dagger \hat{c}_q|\Psi_n\rangle\phi^*_p(\boldsymbol{r})\phi_q(\boldsymbol{r}) - \sum\limits_{p\in \text{HOMOs}} \left|\phi_p(\boldsymbol{r})\right|^2,\nonumber\\
    \rho_\text{e}(t_p)&= 2\Re\left(\sum\limits_{m,n\ge m} C_m^* C_n e^{i(\varepsilon_m-\varepsilon_n)t_p} \sum\limits_{p,q\in \text{LUMOs}}\langle\Psi_m|\hat{c}_p^\dagger \hat{c}_q|\Psi_n\rangle\phi^*_p(\boldsymbol{r})\phi_q(\boldsymbol{r})\right), \label{eq:termElectron}\\
   \rho_\text{h}(t_p)&= 2\Re\left( \sum\limits_{m,n>m} C_m^* C_n e^{i(\varepsilon_m-\varepsilon_n)t_p} \sum\limits_{p,q\in \text{HOMOs}}\langle\Psi_m|\hat{c}_p^\dagger \hat{c}_q|\Psi_n\rangle\phi^*_p(\boldsymbol{r})\phi_q(\boldsymbol{r})\right)\label{eq:termHole}.
\end{align}
\end{widetext}

\bibliography{apssamp.bib}% Produces the bibliography via BibTeX.

\end{document}

\title{Supplemental Material for ''Attosecond momentum-resolved resonant inelastic x-ray scattering for imaging coupled electron-hole dynamics''}
%\\Tracking Electron-Hole Dynamics in Molecules}% Force line breaks with \\

\author{Maksim Radionov\textsuperscript{1,2}}
\author{Daria Popova-Gorelova\textsuperscript{1,2}}%
\affiliation{%
\textsuperscript{1} Institute of Physics, Brandenburg University of Technology Cottbus-Senftenberg,
Erich-Weinert-Straße 1, 03046 Cottbus, Germany
}%
\affiliation{%
\textsuperscript{2} I. Institute for Theoretical Physics and Centre for Free-Electron Laser Science,
Universität Hamburg, Notkestr. 9, 22607 Hamburg, Germany
}%

\maketitle

\begin{figure*}[h]
\centering
\includegraphics[width=0.67\linewidth]{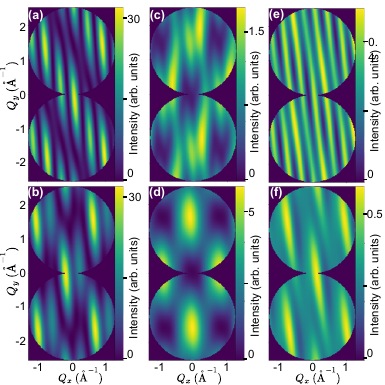}
\caption{The momentum maps $P(Q_x, Q_y, 0)/(\theta(\mathbf n_Q))$ at $Q_z = 0$, with the final state being the ground state at (a) $t_p=1.3$ fs and (b) $t_p=2.7$ fs. (c)--(f) The same as (a)--(b), but with the final state being a valence-excited state at (c) $\omega_\text{s}-\omega_\text{in}=\Delta\omega_{\text{s}2}$ and $t_p=1.3$ fs; (d) $\omega_\text{s}-\omega_\text{in}=\Delta\omega_{\text{s}2}$ and $t_p=2.7$ fs; (e) $\omega_\text{s}-\omega_\text{in}=\Delta\omega_{\text{s}1}$ and $t_p=1.3$ fs and (f) $\omega_\text{s}-\omega_\text{in}=\Delta\omega_{\text{s}1}$ and $t_p=2.7$ fs.}
\label{fig:SupplMomMaps}
\end{figure*}

\begin{figure*}[h]
\includegraphics[width=0.45\linewidth]{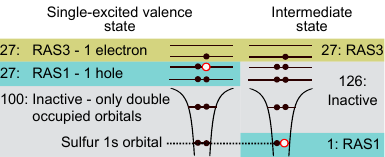}% Here is how to import EPS art
\caption{\label{fig:RASCI}Restricted active spaces for calculations of the valence-excited and 1$s$-excited states.}
\end{figure*}